\documentclass[aps,prb,twocolumn,showpacs]{revtex4}
\usepackage{latexsym}
\usepackage{amssymb}
\usepackage{graphicx}
\begin{document}
\title{A kinetic equation approach to the anomalous Hall effect
in a diffusive Rashba two-dimensional electron system with
magnetization}
\author{S. Y. Liu}
\email{liusy@mail.sjtu.edu.cn}
\author{X. L. Lei}
\affiliation{Department of Physics, Shanghai Jiaotong University, 1954
Huashan Road, Shanghai 200030, China}
\date{\today}

\begin{abstract}

We present a two-band kinetic equation method
to investigate the anomalous Hall effect in a Rashba two-dimensional
electron system subjected to a homogeneous magnetization.
The electron-impurity scattering is taken into account
in the self-consistent Born approximation.
It is demonstrated that the impurity-density-free anomalous Hall
conductivity arises from an
intrinsic and a disorder-mediated mechanisms, associated respectively with the electron states under
and near the Fermi surface. The intrinsic mechanism relates to a
dc-field-induced transition process, or in other words,
a linear stationary Rabi process. The
disorder-mediated one corresponds to a scattering between impurities and
the electrons participating in longitudinal transport.
Numerically, the dependencies of the anomalous Hall conductivity on the spin-orbit coupling constants and
the strength of the magnetization are demonstrated for both the short- and long-range
collisions.

\end{abstract}

\pacs{ 72.20.Fr, 73.50.Dn, 73.63.Hs}
\maketitle

\section{Introduction}

In the presence of a magnetic field, the Hall
effect induced by Lorentz force is a powerful tool for measurement of the
concentration and the nature of the free carriers. However, in many
ferromagnets, a nonvanishing transverse resistivity can also be produced by
the spontaneous magnetization, which exists in the absence of external fields.\cite{AHE}
This so-called anomalous Hall effect (AHE) has been
extensively studied in ferromagnetic semiconductors in recent
years.\cite{Review}

AHE has been first predicted about five decades ago.\cite{Karplus}
At early stages, to interpret it, two mechanisms, namely skew
scattering\cite{Smit} and side-jump process,\cite{Berger} have
been proposed.\cite{Crepieux} These two mechanisms are based on a spin-orbit coupling
included into the potential of electron-impurity
scattering. Obviously, the corresponding contributions to AHE should
rely on the electron-impurity collision.

The recent theoretical interest has been focused on another
"intrinsic" mechanism of
AHE.\cite{Niu1,Niu2,MacDonald1,MacDonald2,
Nagaosa1,Nagaosa2,Kubo1,Kubo2,Kubo3,Kubo4,Sinitsyn} This mechanism has been
first discussed by Karplus and Luttinger,\cite{Karplus} and
clearly rederived by MacDonald {\it et
al}.\cite{MacDonald1,MacDonald2} and Nagaosa {\it et
al}.\cite{Nagaosa1, Nagaosa2} By inclusion of the spin-orbit
interaction into Hamiltonian of the free electrons, nontrivial
transverse conductivity has been obtained even in the absence of
disorder scattering. It makes clear that this AHE is associated with
the Berry phase in the momentum space.\cite{Berry}

More recently, according to the well-known result of St\v reda in
the context of Hall effect in magnetically two-dimensional (2D)
systems,\cite{Streda} Dugaev {\it et al}. discussed another
mechanism of AHE by means of Kubo formula.\cite{Dugaev}
This mechanism is related to electron
states in the vicinity of Fermi surface, and depends on the electron-impurity
scattering but is independent of the impurity density $N_i$.
By considering the short-range disorder collision, it has been found that
this collision-related mechanism makes a contribution to anomalous Hall
conductivity of the same order of magnitude as the intrinsic one.
However, as shown in Ref.\,\onlinecite{Dugaev}, the Kubo approach becomes
questionable when used to investigate the collision-related AHE. Taking the static
limit $\omega\rightarrow 0$ before or after $N_i\rightarrow 0$ leads to
completely different results. At the same time,
up to now, this AHE has been studied only for
short-range electron-impurity scattering. We know that in realistic 2D
semiconductors, the electron density is not large enough to screen the
charged impurities. The Coulomb interaction between electron and impurity
is inevitably long-ranged.

In this paper, we employ a two-band kinetic equation approach to
study the impurity-density-independent AHE in 2D Rashba electron systems with
magnetization. Within this method, we can consider the
time-independent nature of dc transport from the beginning and
hence the problem caused by taking the static limit in Kubo
formalism can be avoided. At the same time,
the long-range electron-impurity collision can be easily handled.
The obtained kinetic equations in the helicity basis allow us to
interpret the above two mechanisms of AHE
in terms of interband processes. The intrinsic AHE
arises from a dc-field-induced direct transition between two
unperturbed spin-orbit-coupled bands. In another point of view,
this AHE can also be understood as a result of a quantum
interference between {\it perturbed} electrons in different bands in
the first order of dc field, {\it i.e.} as a linear stationary Rabi
process. On the other hand, the electrons joining in the longitudinal
transport can be resonantly scattered by the impurities,
leading to another interband polarization independent of impurity
density. This process corresponds to the collision-related mechanism
of AHE. Similar picture has been demonstrated in the
previous studies on the spin-Hall effect in 2D systems with spin-orbit
coupling.\cite{Liu} Based on the derived equations, we numerically
investigate the dependencies of
the AHE on the spin-orbit coupling constant and the strength of
exchange field. At the same time, the comparison between the effects
of short- and long-range disorders on AHE is made.

The paper is organized as follows. In Sec. II we derive the
kinetic equation for nonequilibrium distribution functions and
attempt to resolve it. The above-mentioned two mechanisms of
AHE correspond to two distinct components of the
solution. In Sec. III we perform a numerical calculation to
investigate the effect of spin-orbit coupling constant and magnetization on
AHE. Finally, we
conclude our results in Sec. IV.

\section{Formalism}

\subsection{Kinetic equation}

We consider a quasi-two-dimensional system in the $x-y$ plane
with Rashba spin-orbit interaction. This system is subjected to
a homogeneous magnetization ${\bf M_0}$ along the $z$-direction.
The noninteracting Hamiltonian
of the considered system has the form
\begin{equation}
{\bar H}_0=\varepsilon_{p}+\alpha (\sigma_xp_y-\sigma_y p_x)-M\sigma_z,\label{H0}
\end{equation}
%\end{widetext}
where, ${\bf p}\equiv (p_x,p_y) \equiv (p\cos \phi_{\bf p},p\sin
\phi_{\bf p})$ is the electron momentum, $\varepsilon_p=p^2/2m$,
$\alpha$ is the spin-orbit coupling constant, $M=g\mu_B M_0$, and
$\sigma_i$ ($i=x,y,z$) are the Pauli matrices. This Hamiltonian
can be diagonalized easily, resulting in two eigen wave-functions
$\varphi_{\mu{\bf p}}^{(0)}({\bf r})\equiv u_\mu ({\bf p}){\rm
e}^{i{\bf p}\cdot {\bf r}}$
\begin{equation}
u_\mu ({\bf p})
=\frac{1}{\sqrt{2\lambda_p}}\left (
\begin{array}{c}
\sqrt{\lambda_p-(-1)^{\mu} M}\\
(-1)^{\mu+1}i{\rm e}^{i\phi_{\bf p}}\sqrt{
\lambda_p+(-1)^{\mu}M}
\end{array}
\right ),
\end{equation}
and eigenvalues
\begin{equation}
\varepsilon_\mu (p)=p^2/2m+(-1)^\mu \lambda_p,
\end{equation}
with $\lambda_p\equiv \sqrt{M^2+\alpha^2 p^2}$ and $\mu=1,2$. It
is useful to introduce a unitary transformation $U_{\bf
p}=(u_1({\bf p}),u_2({\bf p}))$, by which the basis of the system
is changed from a spin one to a helicity one. At the same time,
Hamiltonian (\ref{H0}) becomes a diagonal matrix $H_0\equiv
U_{\bf p}^+{\bar H}_0U_{\bf p} ={\rm
diag}(\varepsilon_1(p),\varepsilon_2(p))$.

We consider an electric current flowing along the $x$ axis
when the system is driven by a weak dc field ${\bf E}$
along the $y$-direction. In the helicity basis,
the single-particle operator of this current,
$j_x({ \bf p})\equiv U^+_{\bf p}{\bar j}_x({\bf p})U_{\bf p}$, is given by
\begin{equation}
    j_x({ \bf p})=e\left (
    \begin{array}{cc}
        \frac{p_x}{m\lambda_p}(\lambda_p-m\alpha^2)&
        \frac{\alpha}{\lambda_pp}\left (M
         p_x+i \lambda_p p_y\right )\\
        \frac{\alpha}{\lambda_pp}\left (M
         p_x-i \lambda_p p_y\right )&
        \frac{p_x}{m\lambda_p}(\lambda_p+m\alpha^2)
    \end{array}
    \right ),
\end{equation}
and the corresponding macroscopical quantity is obtained by taking
the statistical average over it, $J_x=\sum_{\bf p}{\rm Tr}[j_x({\bf
p})\rho({\bf p})]$, with $\rho({\bf p})$ being the distribution
function. By definition, the anomalous Hall conductivity is
determined by $\sigma_{xy}=J_x/E$.

In the spin basis, the lesser Green's function, ${\bar {\rm G}}^<$, is a $2\times 2$ matrix and
obeys the Dyson equation
\begin{equation}
    [{\bar {\rm G}}_0^{-1}-U, {\bar {\rm G}}^<]={\bar \Sigma}^r{\bar {\rm G}}^<
    +{\bar \Sigma}^<{\bar {\rm G}}^a-
    {\bar {\rm G}}^r {\bar \Sigma}^<-{\bar {\rm G}}^<{\bar \Sigma}^a, \label{DSON}
\end{equation}
where $U$ is the one-body external potential due to the dc field.
The self-energies ${\bar \Sigma}^{<,a}$ arise from the electron-impurity interaction.
In this paper, we consider a Coulomb interaction between electrons and impurities, which can be described
by a potential $V({\bf p}-{\bf k})$. This potential corresponds to a scattering of an electron
from momentum state ${\bf p}$ to state ${\bf k}$. Under homogeneous condition, Eq.\,(\ref{DSON})
in the momentum space can be written as
\begin{equation}
\left \{\frac{\partial}{\partial T}+e{\bf E}\cdot \nabla_{\bf p}\right \}
{\bar \rho}({\bf p},T)+i[{\bar H}_0,{\bar \rho}({\bf p}, T)]=-\left .
\frac{\partial {\bar\rho}}{\partial T}\right |_{\rm scatt}\label{KE}
\end{equation}
with the Wigner distribution function ${\bar \rho}({\bf p},T)=-i{\bar {\rm G}}^<({\bf p},T,T)$.

We define a distribution function in helicity basis as
$\rho({\bf p},T)=U_{\bf p}^+ {\bar \rho}({\bf p},T)U_{\bf p}$. To
derive the kinetic equation for $\rho$, we multiply the Eq.
(\ref{KE}) from left by $U_{\bf p}^+$ and from right by $U_{\bf
p}$. Due to the unitarity of the transformation, the right-hand side
(rhs) of equation is obtained simply by replacing the Green's
function in spin basis with that one in helicity basis
\begin{equation}
\left . \frac{\partial {\rho}}{\partial T}\right |_{\rm
scatt}=\int_{-\infty}^T dt'[
    \{\Sigma^>,{\rm G}^<\}_+-\{\Sigma^<,{\rm G}^>\}_+](T,t')(t',T).\label{rhs}
\end{equation}
At the same time, using the facts $U_{\bf p}^+\nabla_{\bf p} {\bar
\rho}U_{\bf p}=\nabla_{\bf p}\rho -\nabla_{\bf p}U_{\bf p}^+U_{\bf
p}\rho-\rho U_{\bf p}^+\nabla_{{\bf p}} U_{\bf p}$ and
$\nabla_{\bf p} U_{\bf p}^+U_{\bf p}=-U_{\bf p}^+\nabla_{\bf
p}U_{\bf p}$, we find that the left-hand side (lhs) of the
kinetic equation for distribution function in the helicity basis
takes the form
%\begin{widetext}
\begin{equation}
\left \{\frac{\partial}{\partial T}+e{\bf E}\cdot \nabla_{\bf
p}\right \} \rho-e{\bf E}\cdot [\rho, U_{\bf p}^+\nabla_{\bf p}
U_{\bf p}] +i[H_0,\rho].
\end{equation}
%\end{widetext}

Further, we assume that the applied dc field is weak enough and
only the stationary linear response of the system needs to be considered.
Hence, we can divide the distribution function into two
parts, $\rho=\rho_0+\rho_1$, with
unperturbed distribution function
$\rho_0({\bf p})= {\rm
diag}(n_{\rm F}[\varepsilon_1(p)],n_{\rm F}[\varepsilon_2(p)])$
and the function in the first order of the dc field $\rho_1$.

To simplify relaxation term (\ref{rhs}), it is necessary to
express the two-time lesser and greater Green's functions through
one-time Wigner distribution. The powerful tool to perform this is
the two-band generalized Kadanoff-Baym ansatz (GKBA),\cite{GKBA}
which has been successfully applied to investigate the optical
properties of semiconductors.\cite{Jauho} In the first order of
dc-field strength, the GKBA reads
%\begin{widetext}
\begin{equation}
    {\rm G}^<_1=-{\rm G}_0^r\rho_1+\rho_1{\rm G}_0^a
    -{\rm G}_1^r\rho_0+\rho_0{\rm G}_1^a,\label{GKBA1}
\end{equation}
\begin{equation}
    {\rm G}^>_1=-{\rm G}_0^r\rho_1+\rho_1{\rm G}_0^a
    +{\rm G}_1^r(I-\rho_0)-(I-\rho_0){\rm G}_1^a,\label{GKBA2}
\end{equation}
%\end{widetext}
where, ${\rm G}_0^{r,a}$ and ${\rm G}_1^{r,a}$ are the Green's
functions in the zero- and first-order of dc field,
respectively. Note that ${\rm G}_1^{r,a}$ are off-diagonal matrices. This
fact is obvious from the Dyson equations which they should satisfy.
In Eqs.\,(\ref{GKBA1}) and (\ref{GKBA2}), for shortness, the
time arguments in Green's functions $(t_1,t_2)$ are not written
out. Under the stationary condition, the $\rho_1$ is independent
of the time.

After these simplifications, the scattering term can be formally
denoted as a sum of two components: $\left .\frac{\partial
\rho}{\partial T}\right |_{\rm scatt}\equiv \left .\frac{\partial
\rho}{\partial T}\right |_{\rm scatt}^I+\left .\frac{\partial
\rho}{\partial T}\right |_{\rm scatt}^{II}$. $\left
.\frac{\partial \rho}{\partial T}\right |_{\rm scatt}^{II}$ is off-diagonal
and contains all the elements, associated with the functions ${\rm
G}_1^{r,a}$. In this component, the perturbed
distribution does not appear because
we are only interested in
the linear response of the system.  $\left .\frac{\partial
\rho}{\partial T}\right |_{\rm scatt}^{I}$ includes all the
remaining part.

It is obvious that the driving force in the lhs of kinetic equation, {\it i.e.} the term
proportional to $E$, comprises two
components: the first of which, $e{\bf E}\cdot\nabla_{\bf p}\rho_0$,
contains only diagonal elements, while the another one is
off-diagonal. According to this fact, we break the kinetic equation
into two,
\begin{equation}
e{\bf E}\cdot \nabla_{\bf p}
\rho_0+i[H_0,\rho_1^I]=-\left .\frac{\partial
\rho}{\partial T}\right |_{\rm scatt}^{I},\label{EQ1}
\end{equation}
\begin{equation}
   -e{\bf E}\cdot [\rho_0, U_{\bf p}^+\nabla_{\bf p} U_{\bf p}]
    +i[H_0,\rho_1^{II}]=-\left .\frac{\partial
    \rho}{\partial T}\right |_{\rm scatt}^{II},\label{EQ2}
\end{equation}
which have two solutions $\rho_1^I$ and $\rho_1^{II}$, respectively:
$\rho_1=\rho_1^{I}+\rho_1^{II}$.

In this paper, we restrict ourselves to consider the
AHE being linear in dc field and to the leading order in
the impurity-density expansion. In this case, the Eqs.\,(\ref{EQ1}) and (\ref{EQ2}) are
independent of each other. In fact, from the definition of
 $\left
.\frac{\partial \rho}{\partial T}\right |_{\rm scatt}^{II}$, it is evident that
the perturbed distribution functions $\rho_1^{I}$ and $\rho_1^{II}$ do not
enter in the scattering
term of off-diagonal Eq.\,(\ref{EQ2}). Hence, this equation
can be solved independently. On the other hand, the rhs of Eq.\,(\ref{EQ1}), $\left
.\frac{\partial \rho}{\partial T}\right |_{\rm scatt}^{I}$, generally
depends on the distribution $\rho_1^{II}$.
However, as can be seen below,
the contribution to $\left
.\frac{\partial \rho}{\partial T}\right |_{\rm scatt}^{I}$ from the off-diagonal $\rho_1^{II}$ is
of higher-order of impurity density and can be neglected.
Consequently, the Eqs.\,(\ref{EQ1}) and (\ref{EQ2}) become independent
and can be resolved separately.

We should note that there is a nonvanishing solution of the
Eq.\,(\ref{EQ2}) if the collision term on the rhs is
ignored. This $\rho_1^{II}$ leads to a Hall conductivity independent
of any electron-impurity collision and results in the well-known
intrinsic AHE. We will show below that neglect of the rhs of the
Eq.\,(\ref{EQ2}) is reasonable in the transport study. However,
to resolve Eq.\,(\ref{EQ1}), the scattering term can not be disregarded.
Hence, the anomalous Hall conductivity produced by the corresponding
solution $\rho_1^{I}$ becomes collision-related.

\subsection{Intrinsic anomalous Hall effect}

To carry out the expression of $\rho_1^{II}$, we start with
analyzing the retarded and advanced Green's functions to the first
order of dc field, ${\rm G}_1^{r,a}$,
which appear in the scattering term of the rhs
of Eq.\,(\ref{EQ2}). It is
well known that ${\rm G}_1^{r,a}$ vanish in a one-band electron
gas.\cite{Mahan} From the Dyson equations we can see that in the case
of two bands, the diagonal elements of ${\rm G}_1^{r,a}$ still vanish but their off-diagonal
elements are nonzero and proportional to $({\rm G}_0^{r,a})_{11}-({\rm
G}_0^{r,a})_{22}$, ${\rm G}^{r,a}_1=\frac {i}{2\lambda_p} e{\bf
E}\cdot \sigma_z[{\rm G}_0^{r,a}, U_{\bf p}^+\nabla_{\bf p} U_{\bf
p}]$. Inserting ${\rm G}^{r,a}_1$ into the scattering term, we
obtain
\begin{widetext}
\begin{equation}
    ({\rho}_1^{II})_{\mu {\bar \mu}}({\bf p})=\frac{ie}{\pi}{\bf E}\cdot \nabla_{\bf p}u^+_{\mu}({\bf p})u_{\bar \mu}({\bf p})\int d\omega
    n_{\rm F}(\omega)
    \frac{{\rm Im}[ ({\rm G}^r_0)_{\mu\mu}-({\rm G}^r_0)_{ {\bar\mu}{\bar\mu}}]}
    {\varepsilon_{\mu}(p)-\varepsilon_{ \bar  \mu}(p)},
\end{equation}
\end{widetext}
with ${\bar \mu}=3-\mu$. This expression can be further simplified
if we consider that the collision broadening of the
perturbed Green's functions ${\rm G}_0^{r,a}$
play secondary roles in transport and the imaginary parts of these
functions reduce to $\delta$-functions. In result, the $\rho_1^{II}$ has a
simple form
\begin{equation}
    (\rho_1^{II})_{\mu{\bar \mu}}({\bf p})=
    i e{\bf E}\cdot \nabla_{\bf p}u^+_{\mu}({\bf p})u_{\bar \mu}({\bf p})
    \frac{n_{\rm F}[\varepsilon_{\bar\mu}(p)]-n_{\rm F}[\varepsilon_{\mu}(p)]}
    {\varepsilon_{\mu}(p)-\varepsilon_{ \bar  \mu}(p)}.\label{IR}
\end{equation}
Note that this result can also be derived if we simply neglect the
scattering term on the rhs of Eq.\,(\ref{EQ2}).
Taking the statistical average over the current operator $j_x({\bf
p})$, we find the contribution of $\rho_1^{II}$ to anomalous Hall
conductivity:
\begin{equation}
\sigma_{xy}^{II}=\frac{M\alpha^2 e^2}{2}\int \frac{d{\bf p}}{(2\pi)^2}\frac{1}{\lambda_p^3}
(n_{\rm F}[\varepsilon_1(p)]-n_{\rm F}[\varepsilon_2(p)]).\label{IAHE}
\end{equation}

This result agrees with the previous study on intrinsic anomalous
Hall effect.\cite{Dugaev} Obviously, it is independent of any
electron-impurity scattering and connects with all the electron
states under the Fermi surface. Further, it is clear that this
expression of $\sigma_{xy}^{II}$ is related to the Berry phase
and the topology of the energy bands.\cite{Dugaev} Here, we will
propose another interpretation of this intrinsic
AHE by extending the treatment of Zhang and Yang in the study on
spin-Hall effect.\cite{Zhang}

In fact, the intrinsic contribution to AHE originates from a quantum
interference between two bands perturbed by the external field and
is associated with  the nonvanishing interband dipole moment. The
wavefunction up to the first order of the electric field can be
written as,
\begin{equation}
|\varphi_{\mu{\bf p}}>=|\varphi_{\mu{\bf p}}^{(0)}>+|\varphi_{\mu{\bf p}}^{(1)}>
\end{equation}
with
\begin{equation}
|\varphi_{\mu{\bf p}}^{(1)}>=\sum_{{\bf k}}\frac{
<\varphi^{(0)}_{{\bar \mu}{\bf k}}|
e{\bf E}\cdot {\bf r}|\varphi^{(0)}_{{\mu}{\bf p}}>}
{\varepsilon_{\mu p}-\varepsilon_{{\bar \mu}k}}
|\varphi^{(0)}_{{\bar \mu}{\bf k}}>\label{WF1}
\end{equation}
being the first order perturbation. In order to take into account
the statistical characters of the electrons, we
will utilize the second quantization formalism. The field
operators have the forms, $\psi_{\mu {\bf p}}=|\varphi_{\mu {\bf
p}}^{(0)}>c_{\mu {\bf p}}+ |\varphi_{\mu {\bf
p}}^{(1)}>c_{{\bar\mu} {\bf p}}$ and $\psi_{ {\bf
p}\mu}^+=<\varphi_{\mu {\bf p}}^{(0)}|c^+_{\mu {\bf p}}+
<\varphi_{\mu {\bf p}}^{(1)}|c^+_{{\bar\mu} {\bf p}}$, with the
electron creation and annihilation operators, $c^+_{\mu {\bf p}}$
and $c_{\mu {\bf p}}$. By definition, the interband polarization
$\rho_{\mu{\bar \mu}}({\bf p})$ describes the quantum interference
of {\it perturbed} electrons in different bands,
$\rho_{\mu{\bar \mu}}({\bf p})=<\psi^+_{ {\bf p}{\bar \mu}}
\psi_{ {\bf p}{\mu}}>$.\cite{Jauho} In the first order of dc
field, it reads
\begin{equation}
    (\rho_1^{II})_{\mu{\bar\mu}}({\bf p})=
    <\varphi^{(0)}_{{\bar\mu}{\bf p}}|\varphi^{(1)}_{{\mu}{\bf p}}>
    \{n_{\rm F}[\varepsilon_{\bar \mu}(p)]-n_{\rm F}[\varepsilon_{\mu}(p)]\},
\end{equation}
where the relations
$<\varphi^{(0)}_{{\bar\mu}{\bf p}}|\varphi^{(1)}_{{\mu}{\bf p}}>=
-<\varphi^{(1)}_{{\bar\mu}{\bf p}}|\varphi^{(0)}_{{\mu}{\bf p}}>$ and
$<c^+_{ \mu{\bf p}}c_{\mu {\bf p}}>=n_{\rm F}[\varepsilon_\mu (p)]$ are
used. Substituting Eq. (\ref{WF1}) into this equation
and considering the fact that $<\varphi^{(0)}_{{\bar\mu}{\bf k}}|
e{\bf E}\cdot {\bf r}|\varphi^{(0)}_{{\mu}{\bf p}}>
=iu_{{\bar \mu}{\bf k}}^+e{\bf E}\cdot \nabla_{\bf p}u_{\mu\bf p}\delta({ \bf p}-{\bf k})$,
we finally arrive at expression (\ref{IR}).

It is obvious that this dc-field-induced transition can be
interpreted as a linear stationary Rabi process. The well-known
Rabi oscillation occurs in the presence of ac field and has been
widely investigated from the viewpoints of optics, semiconductors,
and atomic physics {\it etc}. However, the ac field is not the
essential factor of occurring the interference, whereas it ensures
the time oscillation of the Rabi process. If the applied ac field
is replaced by a dc one, the quantum interference also may take place.
For that, the necessary condition is the
nonvanishing of Rabi frequency, which is proportional to the
dipole moment, and the strength of external field.

At the same time, from a perturbative point of view, this
component of interband polarization can also be understood as an
interband transition between two {\it unperturbed} bands. In
this picture, all electrons have finite probability to transit
from one band to another. In result, the nonvanishing
interband polarization emerges and relates to all the
electron states. Note that there is no effect of this transition
on diagonal elements of distribution up to the first order of dc field.

\subsection{Disorder-mediated anomalous Hall effect}

The distribution functions $\rho_1^I$ satisfy the
Eq.\,(\ref{EQ1}). The lhs of this equation consists of
a diagonal driving term $e{\bf E}\cdot \nabla_{\bf p}n_{\rm
F}(\varepsilon_\mu(p))$, and an off-diagonal matrix $i[H_0,\rho_1^{I}]$
depending only on off-diagonal elements of distribution function. Before
seeking the solution of this equation, we should carry out the
self-energy in the self-consistent Born approximation. It takes
a complicated form and is presented in the Appendix. Due to the
spin-orbit coupling, the wave functions possess an additional
momentum dependence, leading to a nonglobal transformation. In result,
each element of the self-energy, and hence the scattering term in
Eq.\,(\ref{EQ1}), becomes a function of all the elements of matrix Green's
function. This fact indicates that the
scattering can induce an admixture of the spin-orbit-coupled
bands as well as the external field. In result, some novel processes
contributing to the distribution functions appear.

To simplify the relaxation term standing on the rhs of
Eq.\,(\ref{EQ1}), we first analyze the lowest exponent in the
impurity-density expansion for the elements of distribution function.
Note that the scattering always provides a contribution of order
of $N_i$.
Since the diagonal driving term in the lhs
of Eq.\,(\ref{EQ1}) is independent of impurity density,
the diagonal elements of distribution $(\rho_1^{I})_{\mu\mu}$
should be of order of $(N_i)^{-1}$. We substitute these
$(\rho_1^{I})_{\mu\mu}$ into the off-diagonal parts of the
relaxation term. Since the term $i[H_0,\rho_1^{I}]$ in the lhs of Eq.\,(\ref{EQ1})
is proportional to the off-diagonal elements of the distribution,
the leading order of the
off-diagonal $(\rho_1^I)_{\mu{\bar \mu}}$ in impurity-density expansion should be $(N_i)^0$.
Hence, all terms containing the off-diagonal
$(\rho_1^I)_{\mu{\bar \mu}}$, as well as $(\rho_1^{II})_{\mu{\bar \mu}}$,
in $\left .\frac{\partial
\rho}{\partial T}\right |_{\rm scatt}^{I}$ make contributions
of higher-order of impurity density and therefore can be ignored.
Under these considerations, we find the coupled equations for
diagonal elements of distribution function,
\begin{widetext}
\begin{equation}
-e{\bf E}\cdot \nabla_{\bf p} n_{\rm F}[\varepsilon_\mu(p)]
=\pi \sum_{\bf k} |V({\bf p} -{\bf k})|^2
\{a_1({\bf p},{\bf k})
[(\rho^{I}_1)_{\mu\mu}({\bf p})-(\rho^{I}_1)_{\mu\mu}({\bf k})]
\Delta_{\mu\mu}+
a_2({\bf p},{\bf k})[(\rho^{I}_1)_{\mu\mu}({\bf p})-(\rho_1^{I})_{{\bar
\mu}{\bar \mu}}({\bf k})]\Delta_{\mu{\bar \mu}}\},\label{DDFs}
\end{equation}
\end{widetext}
with $\Delta_{\mu\nu}\equiv
\delta(\varepsilon_\mu({\bf p})-\varepsilon_\nu({\bf k}))$,
$a_1({\bf p},{\bf k})\equiv (\lambda_p\lambda_k+M^2+\alpha^2kp\cos (\phi_{\bf p}-\phi_{\bf k}))/
\lambda_p\lambda_k
$ and
$a_2({\bf p},{\bf k})\equiv(\lambda_p\lambda_k-M^2-\alpha^2kp\cos (\phi_{\bf p}-\phi_{\bf k}))
/\lambda_p\lambda_k$.

The dc field is assumed to be applied along the $y$ axis. Hence, the lhs
of Eq.\,(\ref{DDFs}) depends on the angle of momentum
through a sine function. From the fundamental triangle relation
$\sin\phi_{\bf k}=-\sin(\phi_{\bf p}-\phi_{\bf k})\cos\phi_{\bf
p}+ \cos(\phi_{\bf p}-\phi_{\bf k})\sin\phi_{\bf p}$ and the
symmetry argument that the terms with $\sin(\phi_{\bf p}-\phi_{\bf
k})$ vanish, it can be followed that the diagonal distribution has
a simple angle-dependence
$(\rho_1^I)_{\mu\mu}=eE\Phi_\mu(p)\sin\phi_{\bf p}$. Due to the elastic
nature of electron-impurity scattering, functions
$\Phi_\mu(p)$ can be carried out analytically
\begin{widetext}
\begin{equation}
    \Phi_\mu (p)=-\frac{\partial n_{\rm F}(\varepsilon_\mu(p))}{\partial \varepsilon_\mu(p)}
        \frac{\left (\tau_{1{\bar \mu}{\bar \mu}}^{-1}
    +\tau_{2{\bar \mu} \mu}^{-1}\right )\frac{\partial \varepsilon_{\mu}(p)}{\partial p}
    +\tau_{3\mu{\bar \mu}}^{-1} \frac{\partial \varepsilon_{ \bar \mu}({\tilde p}_\mu)}
    { \partial {\tilde p}_\mu}}
    {\left (\tau_{1{\bar \mu}{\bar \mu}}^{-1}
    +\tau_{2{\bar \mu} \mu}^{-1}\right )
    \left (\tau_{1{\mu}{\mu}}^{-1}
    +\tau_{2\mu{\bar \mu}}^{-1}\right )-\tau_{3{\bar \mu}{\mu}}^{-1}
    \tau_{3\mu{\bar \mu}}^{-1}},\label{DDF1}
\end{equation}
\end{widetext}
where ${\tilde p}_\mu$ is determined from equation, $\varepsilon_{\bar \mu}({\tilde p}_\mu)=
\varepsilon_\mu (p)$, and the different relaxation times $\tau_{i\mu\nu}$ ($i=1..3,\mu,\nu=1,2$)
are defined by
\begin{equation}
\frac {1}{\tau_{i\mu\nu}}=2\pi n_i\sum_k |V({\bf p}-{\bf k})|^2
\Lambda_{i\mu\nu} ({\bf p},{\bf k}),
\end{equation}
with
$\Lambda_{1\mu\nu}({\bf p},{\bf k})=\frac 12(1-\cos (\phi_{ \bf p}-\phi_{\bf k}))a_1({\bf p},{\bf k})\Delta_{\mu\nu}$,
$\Lambda_{2\mu\nu}({\bf p},{\bf k})=\frac 12 a_2({\bf p},{\bf k}) \Delta_{\mu\nu}$
and $\Lambda_{3\mu\nu}({\bf p}-{\bf k})=\frac 12 \cos (\phi_{ \bf p}-\phi_{\bf k})
a_2({\bf p},{\bf k}) \Delta_{\mu\nu}$.

From Eq.\,(\ref{DDFs}), we find that the terms in the rhs
correspond two distinct scattering processes: the first two terms,
{\it i.e.} the terms associated with $\Delta_{\mu\mu}$, describe
the well-known intraband process, while the remaining come from an
interband transition. It should be noted that the latter interband
process entirely arises from the spin-orbit coupling: it
results in a collision-related admixture of two bands. In the
absence of spin-orbit interaction, as shown in the previous
studies, $\rho_{ {\bar \mu}{\bar \mu}}$, at any time, does not
appear in the equations for $\rho_{\mu\mu}$, even when the
external field is strong.\cite{Jauho}

At the same time, we should note that
$\tau_{1\mu\mu}$ is the longitudinal relaxation time defined
in the framework of Bloch equations. However, the $\tau_{2\mu\nu}$ and
$\tau_{3\mu\nu}$ are novel relaxation times
and, up to now, have not been studied in the literature.

Further, the off-diagonal elements of distribution are given by
\begin{widetext}
\begin{equation}
(\rho_1^{I})_{12}=(\rho_1^{I})_{21}^*=\frac{\pi}{4\lambda_p}\sum_{{\bf k}\,\mu=1,2}|V({\bf
p}-{\bf k})|^2a_3({\bf p},{\bf k})(-1)^{\mu}
\{\Delta_{\mu\mu}[(\rho_1^{I})_{\mu\mu}({\bf
p})-(\rho_1^{I})_{\mu\mu}({\bf k})]-\Delta_{\mu{\bar \mu}}[(\rho_1^{I})_{\mu
\mu}({\bf p})-(\rho_1^{I})_{{\bar \mu}{\bar \mu}}({\bf k})]\}\label{NDF}
\end{equation}
%\end{widetext}
with $a_3({\bf p},{\bf k})\equiv [\alpha k\lambda_{p}\sin (\phi_{\bf p}-\phi_{\bf k})
+i\alpha M(p- k \cos(\phi_{\bf p}-\phi_{\bf k}))]/\lambda_k\lambda_p$.
It is obvious that this function depends
on the momentum angle not only through a sine, but also through a cosine function
\begin{equation}
    (\rho_1^I)_{12}({\bf p})=i\zeta_1(p)\sin\phi_{\bf p}+\zeta_2(p)\cos\phi_{\bf p}.\label{NDF1}
\end{equation}
$\zeta_i(p)$ ($i=1,2$) are determined by
%\begin{widetext}
\begin{equation}
    \zeta_1(p)=\frac{eE\pi}{4\lambda_p}\sum_{{\bf k}\,\mu=1,2}|V({\bf
p}-{\bf k})|^2a_4({\bf p},{\bf k})(-1)^{\mu}
\{\Delta_{\mu\mu}\Phi_\mu(p)[1-\cos (\phi_{ \bf p}-\phi_{\bf k})]-\Delta_{\mu{\bar \mu}}
[\Phi_\mu(p)-\Phi_{\bar \mu}({\tilde p}_\mu)\cos (\phi_{ \bf p}-\phi_{\bf k})]\},\label{a1}
\end{equation}
\begin{equation}
    \zeta_2(p)=\frac{eE\pi}{4\lambda_p}\sum_{{\bf k}\,\mu=1,2}|V({\bf
p}-{\bf k})|^2a_5({\bf p},{\bf k})(-1)^{\mu}
\{\Delta_{\mu\mu}\Phi_\mu(p)\sin (\phi_{ \bf p}-\phi_{\bf k})-\Delta_{\mu{\bar \mu}}
\Phi_{\bar \mu}({\tilde p}_\mu)\sin (\phi_{ \bf p}-\phi_{\bf k})\}\label{a2}
\end{equation}
\end{widetext}
with $a_4({\bf p},{\bf k})\equiv \alpha M(p- k \cos(\phi_{\bf p}-\phi_{\bf k}))]
/\lambda_k\lambda_p$ and $a_5({\bf p},{\bf k})=a_3({\bf p},{\bf k})-ia_4({\bf p},{\bf k})$.

Similarly, from Eq.\,(\ref{NDF}) we can see there are two
scattering processes to form the nonvanishing interband
polarization. One of which, associated with the terms proportional
to $\Delta_{\mu{\bar \mu}}$, is the conventional interband
transition of the perturbed electrons. Due to the scattering
admixture of spin-orbit-coupled bands, even an intraband
transition, connecting with the first two terms of expressions (\ref{a1}) and
(\ref{a2}), makes a
contribution to the interband polarization.

Based on these arguments about the scattering processes in
Eqs.\,(\ref{DDFs}) and (\ref{NDF}), we can generalize our results
to the case of arbitrary two-band 2D system. In a two-band system
with noninteracting wave function $u_\mu({\bf p}) {\rm e}^{i{\bf
p}\cdot {\bf r}}$, the interband or intraband processes, induced
by impurity scattering, are described through $u_\mu^+
({\bf p})V({\bf p}-{\bf k})u_\nu({\bf k})$. If the off-diagonal
distribution functions in the scattering term can be omitted,  by
means of the well-known Fermi golden rules, the $\left
.\frac{\partial \rho}{\partial T}\right |_{\rm scatt}^{I}$ can be
written as
\begin{widetext}
\begin{eqnarray}
    \left .\frac{\partial \rho}{\partial T}\right |_{ {\rm scatt},\mu\nu}^{I}&=&
    \pi\sum_{\bf k \sigma}|V({\bf p}-{\bf k})|^2
    u_\mu^+ ({\bf p})u_\sigma({\bf k})u_\sigma^+({\bf k})u_\nu({\bf p})\nonumber\\
&&  \times\left \{
    \Delta_{\mu\sigma}[(\rho_1^I)_{\mu\mu}({\bf p})-(\rho_1^I)_{\sigma\sigma}({\bf k})]
    +\Delta_{\sigma\nu}[(\rho_1^I)_{\nu\nu}({\bf p})-(\rho_1^I)_{\sigma\sigma}({\bf k})]
    \right \}.\label{SCA}
\end{eqnarray}
\end{widetext}
Note that this equation holds for a general two-band 2D system.

After distribution functions are obtained and the statistical
average over $j_x$ is taken, the anomalous Hall current, related
to the longitudinal transport, can be determined by
\begin{equation}
    J_x^I=2\sum_{\bf p}\left [\frac{\alpha M\zeta_2(p)}{\lambda_p}\cos^2\phi_{\bf p}+\alpha
    \zeta_1(p)\sin^2\phi_{\bf p}\right ].\label{DAHE}
\end{equation}
Since the diagonal elements of the current operator is proportional to
$p_x$, the contribution of diagonal distributions to anomalous
Hall current vanishes. It is obvious that such anomalous Hall
current has a magnitude of order of $(N_i)^0$.

The procedure for deriving the off-diagonal distribution tells us
that this collision-related contribution to anomalous Hall effect
originates from a disorder-mediated process. The electrons,
influenced by external dc field, experience the impurity collision
and join in longitudinal transport. Such scattering produces
diagonal elements of distribution function of order of $(N_i)^{-1}$. At the
same time, these electrons scattered by impurities
give rise to an interband polarization of order of $(N_i)^0$.
The disorder only plays an intermediate role. Note
that at each stage of this disorder-mediated process, both
interband and intraband scatterings occur.

Note that this disorder-mediated mechanism and the well-studied
side-jump mechanism\cite{Berger} are identical physically.
Contributions to Hall conductivity from both the mechanisms are
independent of the impurity density but rely on the
electron-disorder collision. At the same time, they all are
related to the electron states near the Fermi surface. However,
formally, these two mechanisms are completely different. It is
well known that the side-jump process corresponds to a lateral
displacement of the center of the wave-packet during the
scattering when the spin-orbit interaction is included into the
potential of the electron-impurity scattering. It is associated
with the scattering-related term in the current
operator.\cite{Crepieux} However, in our study for Rashba
spin-orbit coupling, the current operator becomes
impurity-independent.

\section{Numerical results}

\begin{figure}
\includegraphics [width=0.45\textwidth,clip] {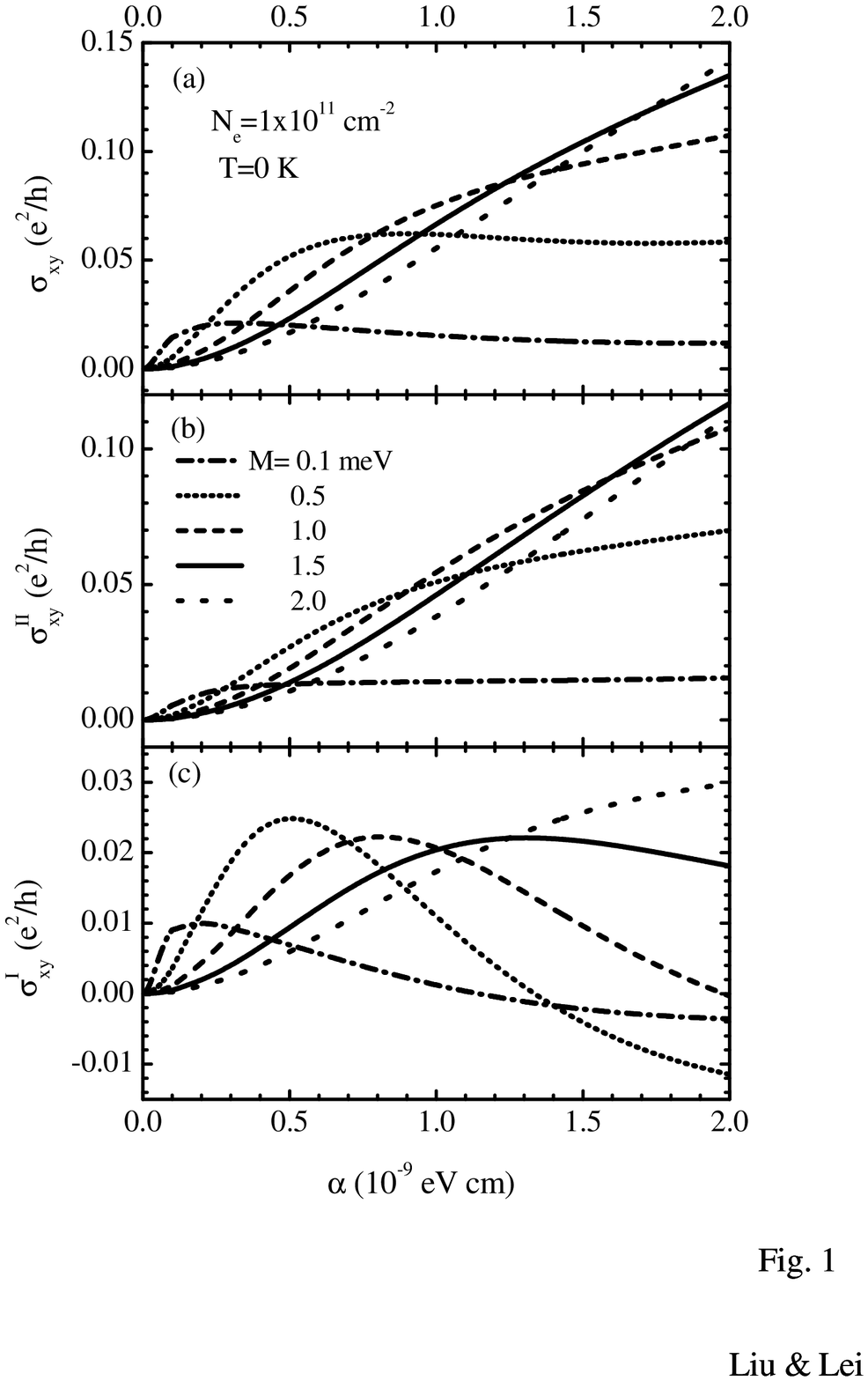}
\caption{Dependencies of disorder-mediated
$\sigma_{xy}^{I}$, intrinsic $\sigma_{xy}^{II}$ and total
anomalous Hall conductivity $\sigma_{xy}$ on the spin-orbit
coupling constant in a Rashba 2D electron
system with magnetization. The lattice temperature $T=0$\,K and
the electron density $N_{\rm e}=1\times 10^{11}$\,cm$^{-2}$. }
\label{fig1}
\end{figure}

\begin{figure}
\includegraphics [width=0.45\textwidth,clip] {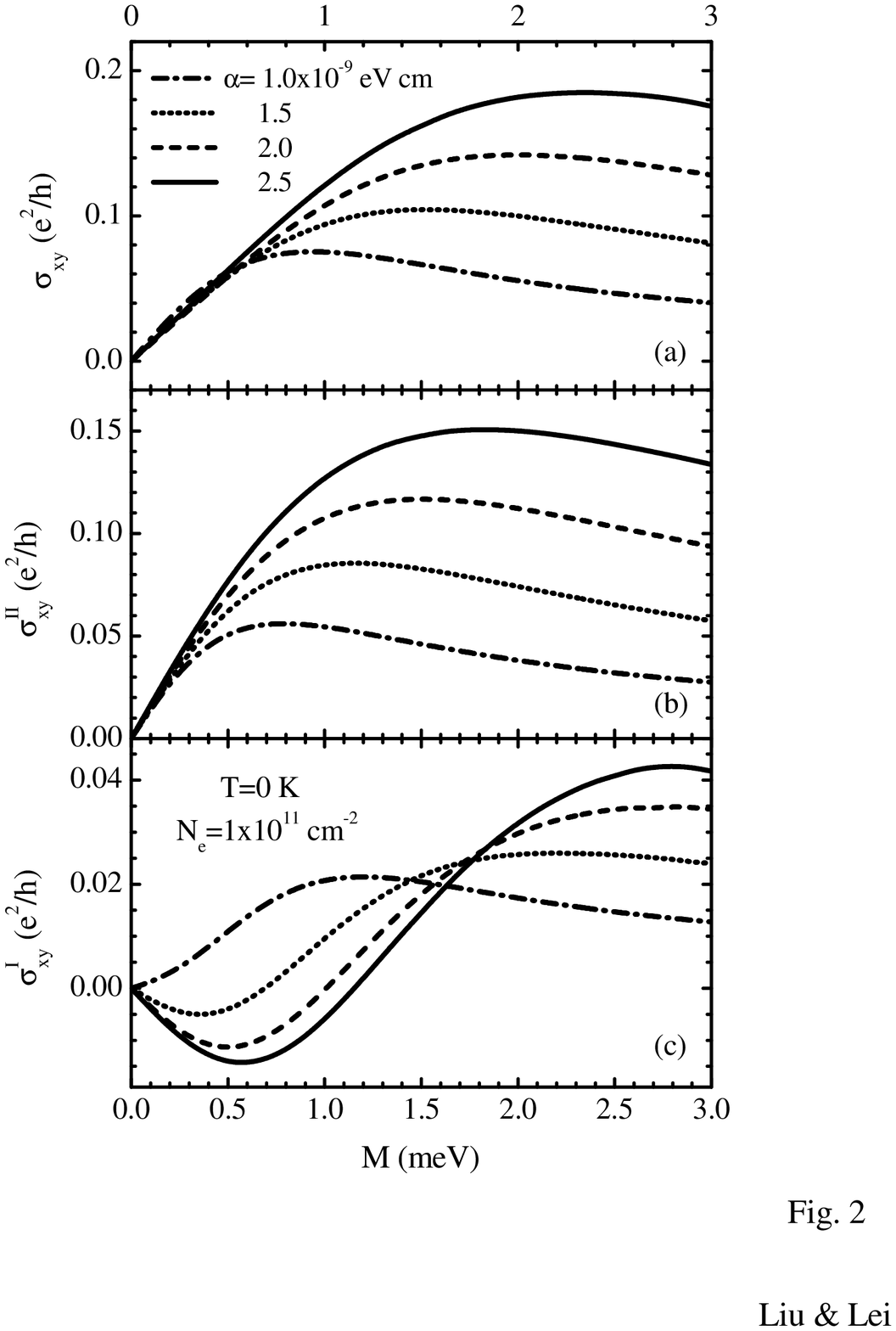}
\caption{Disorder-mediated $\sigma_{xy}^{I}$, intrinsic $\sigma_{xy}^{II}$
 and total anomalous Hall conductivity
$\sigma_{xy}$
as functions of the magnetization. The other parameters are the same as that
in the Fig.\,1.}
\label{fig2}
\end{figure}

\begin{figure}
\includegraphics[width=0.45\textwidth,clip]{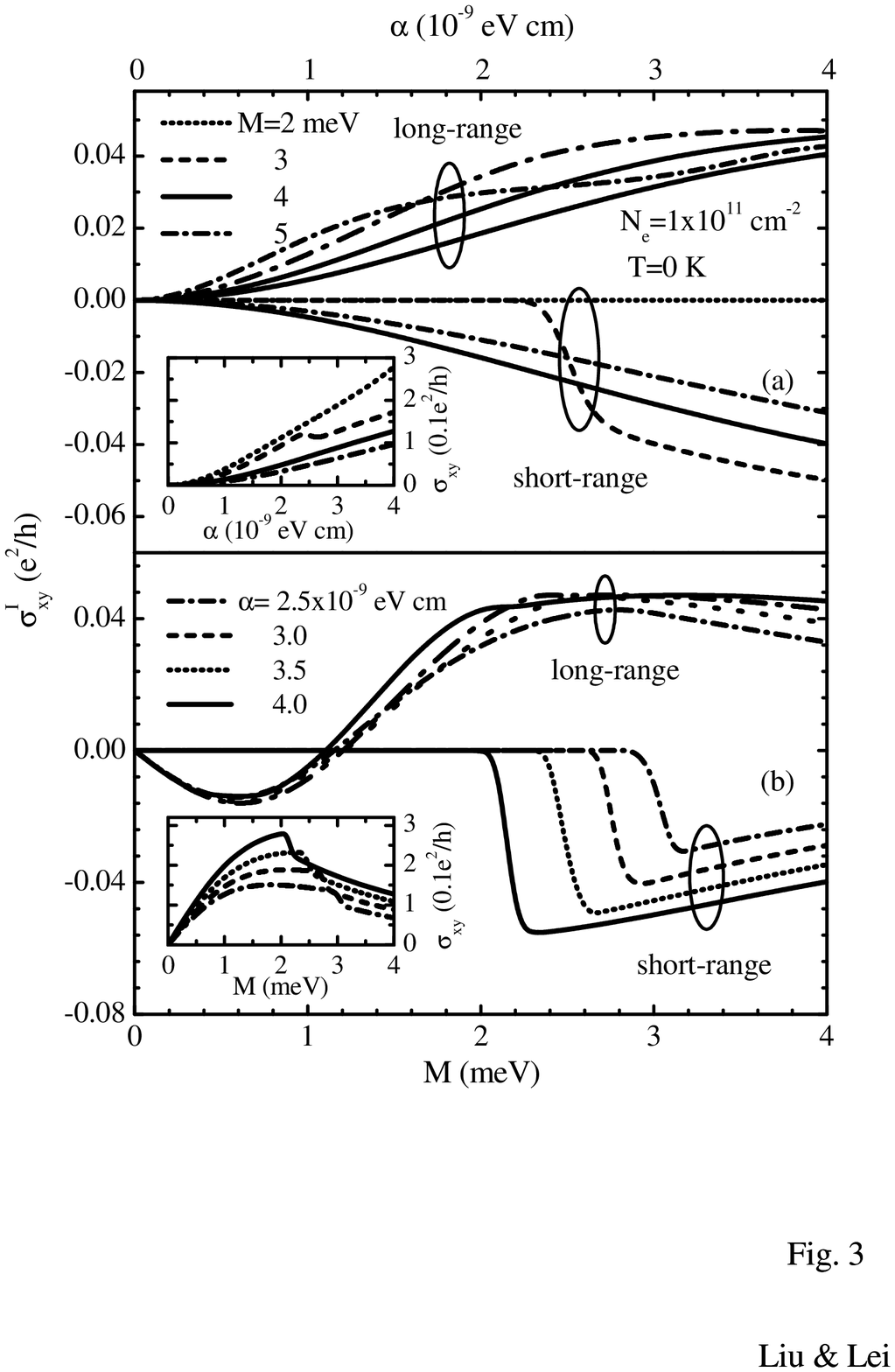}
\caption{Different effects of short- and long-range disorders on the anomalous Hall conductivity.
The coupling-constant and magnetization dependencies of total $\sigma_{xy}$ for short-range collision
are plotted respectively in the insets of Fig.\,3(a) and Fig.\,3(b).}
\end{figure}

We also perform a numerical calculation for a
GaAs/AlGaAs heterojunction
to investigate the intrinsic and disorder-mediated anomalous Hall
effect in a Rashba two-dimensional electron system with
magnetization. In calculation, we first consider a long-range
scattering between electrons and the remote impurities separated
at a distance $s$. The corresponding potential takes the form:
$V(p)=U(p)/\kappa(p)$ with
$U(p)=\frac{e^2}{2\varepsilon_0\kappa p}{\rm e}^{-sp}I(p)$.\cite{Lei}
$I(p)=b^3(b+p)^{-3}$ is the form factor. The wave-function parameter $b$ is given by
$b^3=\frac{33 m e^2}{8\varepsilon_0 \kappa} (N_{\rm e}+\frac{32}{11}N_{{\rm dep}})$. $N_{\rm e}$
and $N_{{\rm dep}}$ are the electron density and the density of
depletion layer charges, respectively. $\kappa(p)=1+q_s H(p)/p$ is the factor coming from the Coulomb screening.
$q_s=\frac{me^2}{2\pi \varepsilon_0 \kappa}$ and $H(p)=\frac 18(8b^3+9pb^2+3bq^2)(b+q)^{-3}$.
In calculation, we take
$m=0.067m_{\rm e}$,  $s=500$\,\AA\,\, and
$N_{{\rm dep}}=1\times 10^{13}$\,m$^{-2}$.
In our formalism, the intrinsic anomalous
Hall effect can be obtained directly by evaluating integral
(\ref{IAHE}). To carry out the disorder-mediated anomalous Hall
current by Eq. (\ref{DAHE}), we should determine first the
diagonal elements of the distribution from Eqs.\,(\ref{DDFs}) and
(\ref{DDF1}) and then the off-diagonal ones by Eq.\,(\ref{NDF1}).
For long-range disorders, the results are plotted in Fig.\,1 and 2.

In Fig.\,1, we plot the disorder-mediated $\sigma_{xy}^{I}$,
intrinsic $\sigma_{xy}^{II}$ and total anomalous Hall conductivity
$\sigma_{xy}$ as functions of the spin-orbit coupling constants. It can be
seen that the values of $\sigma_{xy}^{I}$ and $\sigma_{xy}^{II}$
are comparable. For different magnetization $M$, with increasing
the SO coupling constant, the intrinsic $\sigma_{xy}^{II}$ always
increases, while the $\sigma_{xy}^I$ exhibits a complicated
behavior: for large $M$ it always increases as well as
$\sigma_{xy}^{II}$, but for small $M$ it first increases and then
falls, and even becomes negative for large $\alpha$. The similar
behaviors can also be seen for total $\sigma_{xy}$. It is
clear that $\sigma_{xy}^I$ and $\sigma_{xy}^{II}$ do not always
have the same sign and a compensation occurs for small $M$ and
large $\alpha$.

We also
calculate the dependence of anomalous Hall effect on the
magnetization for different SO coupling constants. The results are shown in
Fig.\,2. Again, we see a sign change for small $M$
and large $\alpha$. Besides, with increasing the magnetization
above a critical value, the $\sigma_{xy}^{I}$ falls. Taking one
with another, in the studied parameter range, the absolute values
of disorder-mediated $\sigma_{xy}^I$ are always lesser than the
intrinsic $\sigma_{xy}^{II}$.

Further, to demonstrate the collision-related feature of the
$\sigma_{xy}^{I}$, we also compute the Hall conductivity for
short-range collision by replacing the potential $V({\bf p}-{\bf k})$
by a momentum-independent one $u$. Note that the resultant
$\sigma_{xy}^{I}$ is independent of the magnitude of $u$. The
calculated results are plotted in Fig.\,3. It can be seen that the
$\sigma_{xy}^{I}$ from the short- and long-range collisions exhibit
completely different behaviors in the dependencies of the SO
coupling constant and magnetization. In the parameter regime:
$\alpha <2\times 10^{-9}$\,eV\,cm and $M<3$\,meV, where the
long-range disorders have strong effect on $\sigma_{xy}$, the Hall
conductivity practically can not be affected by the short-range
disorder. At the same time, in Fig.\,3(b) an abrupt decrease of
$\sigma_{xy}^I$ with increasing the $M$ is seen. Both these features
can be understood from the fact that for short-range collision, the
disorder-mediated mechanism has effect on Hall conductivity approximately
when $M$ becomes larger than the chemical potential. Note that this fact agrees with
the study in Ref.\,\onlinecite{Dugaev}. In Fig.\,3(a),
the $\sigma_{xy}^{I}$ gradually descends with ascending $\alpha$ for
large $M$ since these $M$ are always larger than the chemical
potential in the studied regime of $\alpha$. In contrast to the case
of long-range scattering, disorder-mediated contribution
induced by the short-range collision
always has an opposite sign
with respect to the intrinsic one.

We should note that in the discussion above, to exhibit the effect of different
mechanisms on AHE, we divide the contribution to Hall conductivity into
an intrinsic one and a disorder-mediated one. Actually, in experiment, both
can not be distinguished. They together result in a measurable Hall resistivity
proportional to the square of the impurity density.

\section{Conclusion}
We have proposed a two-band kinetic equation approach to
the anomalous Hall effect in a 2D electron system with
a Rashba spin-orbit interaction and an exchange-field-induced
magnetization. The obtained equation has been resolved by considering
the electron-impurity collision in the self-consistent Born
approximation. It is clear that there exist two contributions
to the impurity-density-free AHE. One of which arises from a
dc-field-induced transition between two spin-orbit-coupled bands,
which can also be understood as a linear stationary Rabi process.
Another contribution is related to the collision, but is independent
of the impurity density. It comes from a process mediated by disorder:
electrons participating in longitudinal transport are scattered
again by impurity, yielding a nonvanishing interband polarization.
Numerically, we have demonstrated the dependencies of both these contributions
on the magnetization and SO coupling-constant : they
always exhibit a compensation for short-range disorder,
while such compensation occurs in the case of long-range collision
only for small magnetization and large SO coupling constants.

Since obtained Eqs.\,({\ref{IR}) and (\ref{SCA}) are valid
even for a general two-band 2D system, the interband polarization
consisting of such two terms is a universal property of 2D systems
with spin-orbit interaction. In result, the quantities, connecting
with the interband polarization, such as anomalous Hall current,
spin-Hall current {\it etc}. should be formed from two distinct
processes, which are associated with the electron states under and near
the Fermi surface, respectively.

\begin{acknowledgments}
One of the authors (SYL) would like to
thank Dr. J. Sinova for bringing our
attention to this problem and gratefully
acknowledge the invaluable discussions with Drs. M. W. Wu, W. Xu,
W. S. Liu, and Y. Chen. This work was supported by the projects of the National
Science Foundation of China and the Shanghai Municipal Commission of Science
and Technology,
and by the Youth
Scientific Research Startup Founds of SJTU.

\end{acknowledgments}

\appendix*
\section{Self-energies}
In the spin basis, electrons experience a spin-independent
long-range disorder collision, which can be described by a
potential $V({\bf p})$. The self-energies for retarded,
advanced and lesser Green's functions in the self-consistent Born
approximation take the forms
\begin{equation}
    {\bar \Sigma}^{r,a,<}({\bf p})=\sum_{\bf k}|V({\bf p}-{\bf k})|^2
    {\bar {\rm G}}^{r,a,<}({\bf k}).
\end{equation}
When a transformation from spin basis to helicity basis is performed,
the forms of the self-energies $\Sigma^{r,a,<}\equiv U^+_{\bf p}{\bar \Sigma}^{r,a,<}U_{\bf p}$
are changed to
\begin{equation}
    { \Sigma}^{r,a,<}({\bf p})=
    \sum_{\bf k}|V({\bf p}-{\bf k}|^2U^+_{\bf p}U_{\bf k}{\rm G}^{r,a,<}({\bf k})
    U^+_{\bf k}U_{\bf p}.\label{SE}
\end{equation}
It can be seen that, due to the non-global feature of the transformation,
the scalar scattering potential in spin basis becomes a matrix in
helicity basis  $T({\bf p},{\bf k})\equiv U^+_{\bf p} V({\bf
p}-{\bf k}) U_{\bf k}$. Generally, each element of self-energy
matrices contains all the elements of Green's functions.

It is convenient to express the self-energies through a sum of
several matrices. Therefor, we first define several matrices:
diagonal matrices ${\rm B}_{i}$ ($i=1..4$)
\begin{equation}
({\rm B}_1)_{\mu\mu}=\frac 12({\rm G}+\sigma_z{\rm G}\sigma_z)_{\mu\mu}=
{\rm G}_{\mu\mu},
\end{equation}
\begin{equation}
({\rm B}_2)_{\mu\mu}=\frac 12(\sigma_x{\rm G}\sigma_x+\sigma_y{\rm G}\sigma_y)_{\mu\mu}=
{\rm G}_{{\bar \mu}{\bar \mu}},
\end{equation}
\begin{equation}
({\rm B}_3)_{\mu\mu}=\frac 12([\sigma_x,{\rm G}]+\sigma_z[\sigma_x,{\rm G}]\sigma_z)_{\mu\mu}=
{\rm G}_{{\bar \mu}\mu}-{\rm G}_{\mu {\bar \mu}},
\end{equation}
\begin{equation}
({\rm B}_4)_{\mu\mu}=\frac 12 ([\sigma_y, {\rm G}]+\sigma_z[\sigma_y,{\rm G}]\sigma_z)_{\mu\mu}=
(-1)^\mu i({\rm G}_{\mu {\bar \mu}}+{\rm G}_{{\bar \mu}\mu}),
\end{equation}
off-diagonal matrices ${\rm C}_i$ ($i=1..3$)
\begin{equation}
({\rm C}_1)_{\mu{\bar \mu}}=\frac 12({\rm G}-\sigma_z{\rm G}\sigma_z)_{\mu{\bar \mu}}=
{\rm G}_{\mu{\bar \mu}},
\end{equation}
\begin{equation}
({\rm C}_2)_{\mu{\bar \mu}}=\frac 12(\sigma_x{\rm G}\sigma_x-\sigma_y{\rm G}\sigma_y)_{\mu{\bar \mu}}=
{\rm G}_{{\bar \mu}\mu},
\end{equation}
\begin{equation}
({\rm C}_3)_{\mu {\bar \mu}}=\frac 12([\sigma_x,{\rm G}]-\sigma_z[\sigma_x,{\rm G}]\sigma_z)_{\mu{\bar \mu}}=
{\rm G}_{{\bar \mu}{\bar \mu}}-{\rm G}_{\mu\mu},
\end{equation}
and matrices ${\rm D}_i=\sigma_z {\rm C}_i$,
$({\rm D}_i)_{\mu {\bar \mu}}=(-1)^{\mu+1}({\rm C}_i)_{\mu {\bar \mu}}$.
By means of these matrices, we can rewrite the self-energies as
\begin{widetext}
\begin{equation}
\Sigma ({\bf p})=\sum_{{\bf k},i=1..4}|V({\bf p}-{\bf k})|^2b_i {\rm B}_i({\bf k})
+\sum_{{\bf k},i=1..3}|V({\bf p}-{\bf k})|^2[c_i {\rm C}_i({\bf k})+d_i{\rm D}_i({\bf k})].
\end{equation}
$b_i$ ($i=1..4$), $c_i$ and $d_i$ ($i=1..3$) are the factors
depending on both the momenta ${\bf p}$ and ${\bf k}$,
\begin{equation}
    b_i=\frac{1}{2\lambda_p\lambda_k}
    [M^2-(-1)^i\lambda_p\lambda_k+\alpha^2 kp \cos(\phi_{\bf p}-\phi_{\bf k})],i=1,2
\end{equation}
\begin{equation}
    b_3=\frac{i\alpha p}{2\lambda_p}\sin(\phi_{\bf p}-\phi_{\bf k}),\,\,
    b_4=\frac{i\alpha M}{2\lambda_p\lambda_k}[k-p\cos(\phi_{\bf p}-\phi_{\bf k})],
\end{equation}
\begin{equation}
c_j=\frac 1{2\lambda_p\lambda_k}[\alpha^2kp+(M^2+(-1)^j\lambda_p\lambda_k)\cos (\phi_{\bf p}-\phi_{\bf k})], j=1,2
\end{equation}
\begin{equation}
c_3=\frac{i\alpha k}{2\lambda_k}\sin(\phi_{\bf p}-\phi_{\bf k}),
\,\,\,\,
%\end{equation}
%\begin{equation}
d_j=\frac{iM}{2\lambda_k\lambda_p}(\lambda_p-(-1)^j\lambda_k),\,j=1,2
\,\,\,\,
%\end{equation}
%\begin{equation}
d_3=\frac{1}{2\lambda_p\lambda_k}(\alpha kM\cos(\phi_{\bf p}-\phi_{\bf k})-\alpha p M).
\end{equation}

In comparison to the systems without
magnetization,\cite{Liu} the self-energies take more complicated forms
due to the additional dependence of wave functions on the
module of momentum.

\end{widetext}

\end{document}